\documentclass[pra,twocolumn,showpacs,eqsecnum,a4paper]{revtex4}
\usepackage{amsbsy}
\usepackage{amsmath}
\usepackage{amsfonts}
\usepackage{graphicx}

\def\slasha#1{\setbox0=\hbox{$#1$}#1\hskip-\wd0\hbox to\wd0{\hss\sl/\/\hss}}
\def\slashb#1{\setbox0=\hbox{$#1$}#1\hskip-\wd0\dimen0=5pt\advance
       \dimen0 by-\ht0\advance\dimen0 by\dp0\lower0.5\dimen0\hbox
         to\wd0{\hss\sl/\/\hss}}

\begin{document}

\title{Photon number superselection and \\
        the entangled coherent state representation}
\author{Barry C. Sanders}
\affiliation{Department of Physics, Macquarie University, Sydney,
        New South Wales 2109, Australia}
\author{Stephen D. Bartlett}
\affiliation{Department of Physics, Macquarie University, Sydney,
        New South Wales 2109, Australia}
\author{Terry Rudolph}
\affiliation{Bell Laboratories, 600-700 Mountain Avenue, Murray Hill,
  NJ 07974, USA}
\author{Peter L.\ Knight}
\affiliation{Optics Section, Blackett Laboratory, Imperial College
        London SW7 2BZ, United Kingdom}
\date{15 October 2003}

\begin{abstract}
  We introduce the entangled coherent state representation, which
  provides a powerful technique for efficiently and elegantly
  describing and analyzing quantum optics sources and detectors while
  respecting the photon number superselection rule that is satisfied
  by all known quantum optics experiments.  We apply the entangled
  coherent state representation to elucidate and resolve the
  longstanding puzzles of the coherence of a laser output field,
  interference between two number states, and dichotomous
  interpretations of quantum teleportation of coherent states.
\end{abstract}
\pacs{03.67.Hk,42.50.Dv,42.50.Ct,03.65.Ud}
\maketitle

\section{Introduction}

Empirically, quantum optics obeys a photon number superselection rule
(PNSSR) due to the lack of an absolute clock or phase standard at
optical frequencies; electromagnetic field sources such as the
laser~\cite{Scu97}, antibunched light sources~\cite{San02} and the
micromaser~\cite{Var00} can be described by incoherent mixtures of
number states, and photodetection described by projective measurement
in the number state basis.  However, coherence is an integral part of
quantum optics, and the coherent state~\cite{Gla63}, which is a
coherent superposition of number states, explicitly violates this
PNSSR.  Pure Gaussian states, such as coherent states and squeezed
states, are very ``convenient fictions''~\cite{Mol97}.  Despite the
PNSSR, the Gaussian state is often attributed ontological significance
when describing things such as the `physical' laser output
field~\cite{Scu97}, the atomic Bose-Einstein condensate~\cite{Rup95},
local oscillators in homodyne detection~\cite{Yue78}, and
continuous-variable quantum teleportation of coherent
states~\cite{Fur98,Zha03,Bow03}.  The ontological view of Gaussian
states is reinforced by optical homodyne tomography, which claims to
reconstruct these states~\cite{Smi93}.  However, such Gaussian states
only appear through a commitment of the partition ensemble fallacy
whereby the density operator is preferentially decomposed into a
mixture of coherent states~\cite{Mol97,Rud01}.

The reason for the preference shown towards Gaussian states over
number states in quantum optics is the coherent state's usefulness as
a representation in interferometry. The essence of its usefulness is
that a linear mode coupling (as in frequency conversion, polarizing
beam splitters and directional couplers), described by a unitary
transformation that conserves the total number of quanta, will
transform a product of two coherent states to another such product
state~\cite{Gla63}.  This simple relation for linear mode coupling is
responsible for the ease of calculating with coherent states over
alternative representations.

Our aim is to introduce a simple method in quantum optics, which is
elegant both as a calculational tool and as a conceptual framework,
that respects the PNSSR (whereby sources produce incoherent mixtures
of number states, and detectors count photons).  We apply this
technique to the challenges of describing interference by mixing
independent number states~\cite{Jav96}, coherence of a multimode laser
output field~\cite{Enk02}, the role of the local oscillator in
homodyne detection, distillable entanglement versus pure entanglement
for two-mode squeezed light~\cite{Enk02}, and the nature of quantum
teleportation of coherent states~\cite{Rud01,Wis02,Fuj03}.  These
applications demonstrate that our operational approach to quantum
optics respecting the PNSSR can quite simply describe all experiments
traditionally described using optical coherence.

Interferometric calculations with number states are tedious: for
$n$-mode coupling, the matrix elements for the unitary transformation
are given by the SU($n$) Wigner $d$-functions~\cite{Row99,Row01}.
Here we show that these calculations are made simple and easy to
interpret by representing number states as entangled coherent
states~\cite{Mec87,San92,San95,San00}, with the entanglement taking
place over a common phase.  This entangled coherent state approach
enables easy calculations with number state sources by exploiting the
ease of using the coherent state representation.  Moreover the
entanglement is not fragile: whereas one normally regards multipartite
entangled coherent states as fragile and challenging to
construct~\cite{Mun01}, the fragility arises due to decoherence with
respect to the optical environment.  For the entangled coherent states
employed here, a decohering mechanism is described by an environment
that is phase-sensitive and thus would violate the PNSSR obeyed by all
sources and measurements.

We begin by reviewing salient points concerning coherent states,
discussing linear mode coupling, the coherent state representation,
and the nature of the laser as a source obeying the PNSSR. We then use
the techniques introduced to analyze interferometry between
independent number states, homodyne detection, squeezed light sources
and continuous variable quantum teleportation.

\section{Concepts and methods}

\subsection{Coherent states and linear mode coupling}
\label{sec:cohst}

A coherent state $|\alpha\rangle$, $\alpha \in \mathbb{C}$, can be
expressed in terms of the Fock states~$|n)$~\cite{Gla63} as
\begin{equation}
  \label{coherentstate}
  |\alpha\rangle \equiv {\rm e}^{-\bar{n}/2}
        \sum_{n=0}^\infty \sqrt{\frac{\bar{n}^n}{n!}} e^{in\varphi} |n)
        \,,
\end{equation}
where $\alpha$ is expressed in polar coordinates as $\alpha =
\sqrt{\bar{n}} e^{i\varphi}$, with amplitude $\sqrt{\bar n}$ (mean
photon number $\bar n$) and phase~$\varphi$.  This coherent state has
photon number statistics given by the Poisson distribution
\begin{equation}
  \label{dist:Pi}
  \Pi_n(\bar{n}) \equiv \text{e}^{-\bar{n}} \frac{\bar{n}^n}{n!} \, ,
\end{equation}
with mean and variance both equal to $\bar{n}$.  The coherent state is
an eigenstate of the annihilation operator~$\hat{a}$, satisfying the
eigenvalue relation
\begin{equation}
    \hat{a}|\alpha\rangle=\alpha|\alpha\rangle.
\end{equation}
It is also a minimum uncertainty state with respect to its conjugate
quadrature operators $\hat{q}\equiv \hat{q}_0$ and
$\hat{p}\equiv\hat{q}_{\pi/2}$ where (choosing units such that
$\hbar\equiv 1$)
\begin{equation}
\label{operator:quadrature}
  \hat{q}_\theta \equiv
    \frac{1}{\sqrt{2}}\left(\text{e}^{\text{i}\theta} \hat{a}
    + \text{e}^{-\text{i}\theta} \hat{a}^{\dagger} \right) \, .
\end{equation}
The canonically conjugate operators satisfy the commutator relation
$[\hat{q},\hat{p}]=\text{i}\openone$, and the minimum uncertainty
relation is thus $\Delta q \Delta p = 1/2$.  The coherent state is a
displaced vacuum state, $|\alpha\rangle=D(\alpha)|0\rangle$, for
$D(\alpha)\equiv\exp(\alpha\hat{a}^\dagger-\alpha^*\hat{a})$.

The properties discussed above are often cited as the key
properties of the coherent state, but another property is crucial
for calculations in quantum optics.  So far we have considered
single-mode coherent states; we introduce the two-mode coherent
state $|\alpha,\beta\rangle\equiv |\alpha\rangle_a \otimes
|\beta\rangle_b$, where $a,b$ label the two modes.  The
Hamiltonian that generates linear mode coupling is given by
\begin{equation}
\label{H:linearcoupling}
        \hat{H} = {\rm i}(g^*
        \hat{a}^\dagger\hat{b}-g\hat{a}\hat{b}^\dagger),
\end{equation}
with $|g|$ quantifying the coupling strength between the two modes and
arg$(g)$ the relative phase shift between the modes imposed by the
coupling.  The Hamiltonian (\ref{H:linearcoupling}) generates the
unitary evolution operator
\begin{equation}
\label{unitary:linearcoupling}
  U(\theta,\phi)= \exp (-{\rm i} \hat{H} t) = \exp\left(\theta
  e^{-{\rm i}\phi}\hat{a}^\dagger\hat{b}
      -\theta e^{{\rm i}\phi} \hat{a}\hat{b}^\dagger\right)
\end{equation}
for $\theta=|g|t$, $\phi={\rm arg}(g)$ and~$t$ the interaction
time.

As is well known, the linear coupling unitary
transformation~(\ref{unitary:linearcoupling}) transforms a two-mode
product coherent state to a two-mode product coherent
state~\cite{Gla63}. The easiest way to establish this property is
first to note that the annihilation operators transform according to
\begin{equation}
\label{operators:linearcoupling}
        U^\dagger(\theta,\phi)
        \begin{pmatrix}\hat{a}\\ \hat{b}\end{pmatrix}U(\theta,\phi)
        =\mathcal{M}(\theta,\phi)
        \begin{pmatrix} \hat{a}\\ \hat{b} \end{pmatrix}
\end{equation}
for
\begin{equation}
    \label{M:theta}
        \mathcal{M}(\theta,\phi) \equiv \begin{pmatrix} \cos\theta
        & \text{e}^{-\text{i}\phi}\sin\theta \\
        -\text{e}^{\text{i}\phi}\sin\theta
        & \cos\theta\end{pmatrix} \, .
\end{equation}
If the input state is the two-mode coherent state, the output is the
eigenstate of the output annihilation operators in
(\ref{operators:linearcoupling}), namely the two-mode coherent state
\begin{multline}
  \label{state:linearcoupling}
  U(\theta,\phi)|\alpha,\beta\rangle \\
  = |\alpha\cos\theta+\beta\text{e}^{-\text{i}\phi}\sin\theta,
     -\alpha\text{e}^{\text{i}\phi}\sin\theta+\beta\cos\theta\rangle \,.
\end{multline}
The condition for 50/50 (or 3 dB) splitting is met if $\theta=\pi/4$.

Another important aspect of coherent states is that they constitute an
overcomplete basis of the Hilbert space for the harmonic oscillator,
giving
\begin{equation}
\int \frac{\text{d}^2\alpha}{2\pi} |\alpha\rangle\langle\alpha|
        = \openone \,,
\end{equation}
with $\openone$ the identity operator.  An arbitrary density operator
can be expressed as
\begin{equation}
\label{rep:P}
     \hat{\rho}=\int \frac{\text{d}^2\alpha}{2\pi} P(\alpha)
     |\alpha\rangle\langle\alpha|,
\end{equation}
with $P(\alpha)$ the Glauber-Sudarshan $P$ representation
\cite{Gla63,Sud63}.  Density operators are said to be nonclassical if
and only if $P(\alpha)$ does not satisfy the axioms of a true
probability density; if it does, the field density operator is
``semiclassical''.

\subsection{Photon number superselection rule}
\label{subsec:PNSSR}

Quantum optics empirically obeys a photon number superselection rule
(PNSSR).  Operationally, a superselection rule can be expressed as an
invariance of all states and operations (unitary transformations,
measurements, dissipation, etc.) by a group action~\cite{Bar03}.  For
a PNSSR, this group is the U(1) group of unitary phase shifts, with
the unitary phase-shift operator given by
\begin{equation}
     \label{unitary:phaseshift}
     \mathcal{P}(\Delta)\equiv\exp(\text{i}\Delta\hat{a}^\dagger\hat{a}),
\end{equation}
$\Delta \in [0,2\pi)$, which transforms the coherent state according
to
\begin{equation}
    \label{coherent:phaseshift}
    \mathcal{P}(\Delta)|\alpha\rangle
    =|\alpha\text{e}^{\text{i}\Delta}\rangle.
\end{equation}
The PNSSR ensures that density operators for quantum optics sources
are U(1) invariant:
\begin{equation}
    \label{rho:invariance}
    \mathcal{P}(\Delta)\hat{\rho}\mathcal{P}^{\dagger}(\Delta) =
    \hat{\rho} \, , \quad \Delta \in \text{U(1)} \, .
\end{equation}
Expressing the integration measure as $\text{d}^2\alpha/\pi =
\slashb{\text{d}}\varphi\text{d}\bar{n}$, where we use the `slash
notation' for the differential operator
$\slashb{\text{d}}\equiv\text{d}/2\pi$, the independence of the
density operator on phase shifts (\ref{rho:invariance}) implies that
$P(\alpha)$ is axisymmetric over the complex-$\alpha$ plane:
\begin{equation}
\label{P:constraint}
     P(\alpha)=P(\sqrt{\bar{n}}).
\end{equation}
This constraint on the representation is quite strong.  The constraint
(\ref{P:constraint}) allows the arbitrary density
operator~(\ref{rep:P}) to be expressed as
\begin{align}
\label{density:axisymmetric}
        \hat{\rho}&=\int_{0}^{2\pi}\slashb{\text{d}}\varphi
        \int_{0}^{\infty} P(\sqrt{\bar{n}})
        |\sqrt{\bar{n}}\text{e}^{\text{i}\varphi}\rangle\langle
        \sqrt{\bar{n}}\text{e}^{\text{i}\varphi} |
                \nonumber       \\
        &=\sum_{n=0}^\infty p_n |n)(n|
\end{align}
with
\begin{equation}
  \label{eq:pn}
     p_n=2\int_0^\infty \text{d}\bar{n}\,\Pi_n(\bar{n})P(\sqrt{\bar n})
\end{equation}
and $\Pi_{n}(\bar{n})$ the Poisson distribution defined by
(\ref{dist:Pi}).

We see that a consequence of the PNSSR is that \emph{any} optical
source can be regarded in two equivalent ways: as a source of coherent
states with quasi-probability distribution $P(\alpha) = P(\sqrt{n})$
that is uniform in phase, or as a source of number states with the
photon number distribution given by~(\ref{eq:pn}).  Each
interpretation is compatible with experimental results; to ascribe
ontological significance to one description over the other is a
fallacy.

\subsection{Entangled coherent state representation}

We have established above that sources satisfying the PNSSR can be
regarded as mixtures of number states.  The challenge of using number
states for interferometric calculations is that matrix elements of the
linear coupling unitary transformation (\ref{unitary:linearcoupling})
in the number state basis are the SU($n$) Wigner $d$ functions, for
example $d^j_{mn}(\theta)$ for two-mode coupling; tools for
efficiently calculating SU(2) and SU(3) Wigner $d$ functions are
available including asymptotic techniques~\cite{Row99,Row01}, but in
the following we establish an easier formalism for quantum optics
calculations that employs a coherent state representation.

The coherent states form a basis, and thus we can represent any number
state in this basis as a superposition of coherent states.  In doing
so, there exists an ambiguity due to the overcompleteness of the
coherent state basis.  Our preference here is to represent the number
state as a superposition of coherent states over a circle in the
complex-$\alpha$ phase space~\cite{Buz95},
\begin{equation}
   \label{state:circle}
   |n) = [\Pi_n(m)]^{-1/2}
   \int \slashb{\text{d}}\varphi \,
   \text{e}^{-\text{i}n\varphi}
   |\sqrt{m}\text{e}^{\text{i}\varphi}\rangle ,
\end{equation}
which is valid for any integer $m>0$.  We choose to fix $m=n$ so
that the number state is presented as a superposition of all
coherent states on the circle with radius $\sqrt{n}$.

The natural extension of Eq.~(\ref{state:circle}) to a two-mode Fock
state is given by
\begin{multline}
  |n,n^{\prime}) = \left[\Pi_n(n) \Pi_{n^\prime}(n^\prime) \right]^{-1/2}
                \\       \times
  \int \slashb{\text{d}}\varphi \slashb{\text{d}}\varphi^\prime
  \text{e}^{-\text{i}(n\varphi +n^\prime\varphi')}
  |\sqrt{n}\text{e}^{\text{i}\varphi},
  \sqrt{n^\prime}\text{e}^{\text{i}\varphi^\prime} \rangle
  \label{entstate:circle}
\end{multline}
with $|\sqrt{n}\text{e}^{\text{i}\varphi},
\sqrt{n^\prime}\text{e}^{\text{i}\varphi^\prime}\rangle$ a two-mode
coherent state.  Although at first glance the right hand side
of~(\ref{entstate:circle}) appears to be a two-mode entangled coherent
state~\cite{Mec87,San92,San95,San00}, it is a product state and hence
not actually entangled.  However, the state becomes a genuine
entangled coherent state subsequent to linear coupling by
(\ref{unitary:linearcoupling}) of the two modes.  The entangled
coherent state representation is a great advantage in studying linear
coupling of number states, as shown in the following.

Consider the linear mode coupling transformation of an input state
consisting of $n$ photons in one mode and no photons (the vacuum state
$|0\rangle$) in the other mode.  In the entangled coherent state
representation we can write
\begin{equation}
\label{input:n}
  |n,0) = \left[\Pi_n(n)\right]^{-1/2}
  \int_{0}^{2\pi}\slashb{\text{d}}\varphi\, \text{e}^{-\text{i}n\varphi}
  |\sqrt{n}e^{\text{i}\varphi},0\rangle .
\end{equation}
The output state, following the transformation
(\ref{unitary:linearcoupling}), is
\begin{align}
\label{output:n}
   U(\theta,\phi)|n,0)
   = & \left[\Pi_n(n)\right]^{-1/2}
   \int_{0}^{2\pi}\slashb{\text{d}}\varphi e^{-\text{i}n\varphi}
        \nonumber       \\ &\times
   |\sqrt{n}\text{e}^{\text{i}\varphi}\cos\theta,
   -\text{e}^{\text{i}(\varphi+\phi)}\sqrt{n}\sin\theta\rangle,
\end{align}
where we have used the results derived in
Eq.~(\ref{state:linearcoupling}).  This output state~(\ref{output:n})
is an entangled coherent state~\cite{Mec87,San92,San95,San00}, with
the entanglement over optical phase; this entanglement is robust
against any decoherence mechanism involving linear coupling to an
environment that also obeys the PNSSR.  Only a decoherence mechanism
that breaks the PNSSR can destroy this entanglement.

The general two-mode Fock state~(\ref{entstate:circle}) transforms
via linear coupling to the entangled coherent state
\begin{align}
\label{output:nn'}
  U(\theta,\phi)|n,n^\prime)
        &=\left[\Pi_n(n)\Pi_{n^\prime}(n^\prime)\right]^{-1/2}
        \int \slashb{\text{d}}\varphi\slashb{\text{d}}\varphi^\prime \,
        \text{e}^{-\text{i}(n\varphi+n'\varphi')}
             \nonumber\\        \times &
        |\sqrt{n}\text{e}^{\text{i}\varphi}\cos\theta
        +\sqrt{n^\prime} e^{\text{i}(\varphi'-\phi)}\sin\theta,
                \nonumber       \\
        &-\sqrt{n}\text{e}^{\text{i}(\varphi+\phi)}\sin\theta
        +\sqrt{n^\prime} e^{\text{i}\varphi'}\cos\theta\rangle ,
\end{align}
with the entanglement over two optical phases $\varphi$ and
$\varphi^\prime$.  Generalization to multimode Fock states is
straightforward.




\section{Sources: the laser field}
\label{sec:number}

An important application of this theory is to the laser output field.
There are standard theories that describe the formation of the
intracavity laser field, which is necessarily diagonal in the number
state representation~\cite{Scu97}. Nevertheless, the field emitted
from the cavity exhibits multimode coherence, and it is tempting to
regard the multimode laser output as being in a multimode coherent
state. A number state in the cavity appears to lead to a highly
entangled multimode output whereas the intracavity coherent state
leads very nicely to a product coherent state in the multimode
extracavity field.

The preference for coherent states is highlighted in a recent
discussion of the ideal laser and its output field by van Enk and
Fuchs (vEF)~\cite{Enk02}.  They express a preference for treating the
laser in terms of coherent states, a view that was originally
championed by Glauber~\cite{Gla63}.  However, the ease of using the
coherent state representation should not be regarded as a
justification for a commitment of the partition ensemble fallacy and
thus regarding number states as less physical.  The formalism
developed here clarifies why a number state in the cavity can equally
well lead to a coherent multimode output.

As the intracavity field is described by an axisymmetric density
matrix of the type~(\ref{density:axisymmetric}), it is equally valid
to describe the source as a distribution of number states or as a
distribution of coherent states.  With the entangled coherent state
formalism, we show that the output field may be regarded as an
entangled coherent state with the entanglement over the optical phase
variable of the laser.  This entangled state can be expressed as a
superposition of product coherent states, which exposes the multimode
coherence of the output field.  However, the reduction to a multimode
coherent state, which is what vEF yearn for in describing their
``complete measurement'' that would collapse the wave function into a
particular overall phase requires that one breaks the PNSSR.  We argue
in the following that there is no need and no justification for
postulating such a decoherence process.  We do not argue that such a
complete measurement is not possible in principle, only that no
process of this type is present in current quantum optics experiments
and would require an absolute clock or phase standard at optical
frequencies.  Without such a complete measurement, the number state
and coherent state sources are equally valid physically, and the
entangled coherent state representation clarifies that a number state
in the cavity produces exactly the desired multimode coherence.

Specifically, the multimode laser output can be described by employing
multiple spectral components, a sequence of pulses, spatial modes or
other possibilities.  The actual nature of the output modes is not
important to this analysis; only the fact that the coupling between
the single-mode intracavity field and the multimode output field is
via a linear coupling mechanism. For simplicity we assume that the
laser is ideal with Poissonian photon statistics according to the
distribution~(\ref{dist:Pi}), and assume that the density operator for
the single-mode field in the laser cavity is
\begin{align}
\label{density:ideallaser}
\hat{\rho}_{\text{L}}(\bar{n})
        &\equiv\int\slashb{\text{d}}\varphi \,
        |\sqrt{\bar{n}}\text{e}^{\text{i}\varphi}\rangle \langle
        \sqrt{\bar{n}}\text{e}^{\text{i}\varphi} |
        \nonumber       \\
        &=\sum_{n=0}^\infty \Pi_n(\bar{n}) |n)(n| ,
\end{align}
which is a mixture of coherent states with amplitude $\sqrt{\bar n}$
in the cavity, uniformly distributed over the optical phase $\varphi$,
and is also a Poissonian mixture of number states with $\bar{n}$ the
mean number of photons.

The laser field output is related to the input field by linear
coupling of the form (\ref{unitary:linearcoupling}), with the
annihilation operator $\hat{b}$ given by a linear combination of
annihilation operators $\hat{b}_{k}$ for each of output field mode.
If we consider, for example, a continuous-wave (cw) output field, the
multimode output is described by a sequence of overlapping pulses
(spread over both time and frequency) that together constitute the
nearly monochromic output field.  This case is the one considered by
vEF.  The appeal of employing coherent states is that the intracavity
state $|\sqrt{\bar n}\text{e}^{\text{i}\varphi}\rangle$ can produce
the $N$-mode product state
\begin{equation}
\label{laser:product}
     |\sqrt{\bar{n}/N}\text{e}^{\text{i}\varphi},\ldots,
     \sqrt{\bar{n}/N}\text{e}^{\text{i}\varphi} \rangle
     =\prod_{k=1}^N
     |\sqrt{\bar{n}/N}\text{e}^{\text{i}\varphi}\rangle_k \, ,
\end{equation}
describing a state for which the photons have been split equally
between the $N$ modes.  The state (\ref{laser:product}) is one
possible description of the laser output field: an initial density
that is diagonal in the number state representation must yield an
output density that is also diagonal in this
representation~\cite{Rud02} unless the PNSSR is broken, which is
certainly not the case for linear coupling.

We now show how a source of number states yields equivalent results.
In analogy to the linear coupling Hamiltonian and initial conditions
that yield the product state~(\ref{laser:product}), we can also
consider a number state $|m)$ in one mode, the vacuum in the other
$N-1$ modes, and the same linear coupling transformation.  The input
state of $m$ photons in the first of $N$ modes and all other modes in
the vacuum state to an equal distribution of photons in all~$N$ modes,
as for (\ref{laser:product}), is given by
\begin{multline}
  \label{output:N}
  \left[\Pi_m(m)\right]^{-\frac{1}{2}}
   \int \slashb{\text{d}}\varphi \text{e}^{\text{i}m\varphi}
   |\sqrt{m} \text{e}^{\text{i}\varphi} ,0,\cdots,0\rangle  \\
   \rightarrow\left[\Pi_m(m)\right]^{-\frac{1}{2}}
   \int \slashb{\text{d}}\varphi \text{e}^{\text{i}m\varphi}
   |\sqrt{m/N}\text{e}^{\text{i}\varphi}, \ldots,
    \sqrt{m/N}\text{e}^{\text{i}\varphi}\rangle .
\end{multline}
This entangled coherent state is a superposition of product coherent
states that are identical in amplitude and phase, with coefficients
of the superposition distributed uniformly over the phase~$\varphi$.

The entangled coherent state represents the output of the laser field
for an $m$-photon number state prepared in the single-mode intracavity
field.  Expression (\ref{output:N}) is \emph{as valid} as expression
(\ref{laser:product}) in describing the output field.  Although the
product coherent state has been championed~\cite{Enk02}, avoiding the
partition ensemble fallacy requires each decomposition to be equally
acceptable.

The laser's coherence time or length can be easily described within
the entangled coherent state representation of the number state
(\ref{output:N}) by including a random walk in the phase.  For the
product coherent state in the following expression representing the
amplitudes of successive overlapping pulses, the ideal laser output
field can be expressed as
\begin{equation}
  \label{output:phasewalk}
  \left[\Pi_m(m)\right]^{-\frac{1}{2}}
   \int \slashb{\text{d}}\varphi \text{e}^{\text{i}m\varphi}
   |\sqrt{m/N}\text{e}^{\text{i}\varphi(t_{1})}, \ldots,
    \sqrt{m/N}\text{e}^{\text{i}\varphi(t_{N})}\rangle ,
\end{equation}
where $\varphi(t)$ is determined for times $\{t_{k}\}$ by a
stochastic sequence.  The sequence can be regarded as a random
walk, and correlations are calculated from the above multimode
entangled coherent state, averaged over all realizations of this
random walk in phase.

With the above expression, the role of vEF's ``complete
measurement''~\cite{Enk02} is clear.  This measurement would ideally
measure the phase of one state either in the product
state~(\ref{laser:product}), or, equivalently, in the entangled
coherent state (\ref{output:N}) and yield a result for the phase. For
the case of the intracavity field described by a number state, the
result is that the entangled coherent state `collapses' to a product
of $n-1$ identical coherent states, regardless of the fact that the
intracavity field initiated as a number state.  Thus, there is no
physical preference for the coherent states as a decomposition of the
density operator.  Moreover, their ``complete measurement'' must break
the PNSSR, which would require an ancilla state such as atoms in a
superposition of different energy eigenstates~\cite{Rud02}.  This
requirement of a superposition of energy eigenstates simply shifts the
burden by allowing phase localization to occur by using a source
wherein phase localization is available.

\section{Interference of number states}

The analysis of interferometry with number states becomes
straightforward when using the entangled coherent state
representation, because an interferometer is a linear mode coupler.
For an interferometer with $N$ modes, the unitary transformations are
elements of the Lie group SU($N-1$)~\cite{San99}. Transformations can
be calculated from matrix elements of the unitary linear coupling
transformation, but the calculations, which involve Wigner $d$
functions, are complicated (although solutions are known for small
$N$~\cite{Row99,Row01}).  The entangled coherent state formalism
offers an elegant alternative.

We now use this formalism to examine the remarkable result that
interference can be observed between independently generated number
states.  Consider an initial state of two modes $a$ and $b$ of the
light field that takes the form of a product of Fock states
$|n_1)_a|n_2)_b$; the modes are subsequently combined at a
beam splitter, followed by photodetection at both output modes.

\begin{figure}
  \includegraphics*[width=3.25in,keepaspectratio]{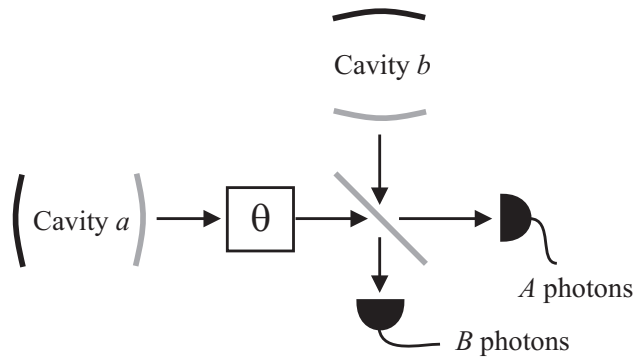}
  \caption{Schematic of a scheme to interfere the output state of two
    cavities at a beam splitter to detect interference.  A partial
    mirror on each cavity (in grey) gives a linear coupling of the
    cavity to the output mode.  These output modes are combined at a
    beam splitter, followed by photodetection.}
  \label{fig:FockInterference}
\end{figure}

It is well accepted that if the initial states of the two modes were
coherent states, then an interference pattern will be recorded at the
two detectors. This interference pattern could be passively observed
as a function of time (if the two modes were at slightly different
frequencies), or as a function of some actively varied phase shift
$\theta$ introduced in one of the modes just prior to the beam splitter
(see Fig.~\ref{fig:FockInterference}).
It is often stated, however, that, since the first order correlation
function $g^{(1)}$ vanishes for the state $|n_1)_a|n_2)_b$, no
interference will be observed in this case.  (``This [mixing of number
states] yields a zero correlation function and thus no fringes are
obtained.'' - Ref.~\cite{Wal94}, p. 38.)  Such arguments are sometimes
then applied to the Pfleegor-Mandel experiments~\cite{Pfl67}, in which
interference patterns are observed between the outputs of two
different lasers, in order to claim that the laser output is
necessarily a coherent state.

As we now show, these arguments are erroneous - they ultimately arise
from a misconception about the role of correlation functions in
determining operationally observable properties of the electromagnetic
field.  M{\o}lmer has shown~\cite{Mol97,Mol97b}, through intensive
calculations and numerical simulations, how two independent Fock
states can interfere.  We employ the entangled coherent state
representation to show this result analytically through a much simpler
analysis.  Although our results are phrased in terms of interference
between photons, they apply equally well to other bosonic modes such
as Bose-Einstein condensates. In fact, by our technique we can
reproduce the celebrated result of Javanainen and Yoo~\cite{Jav96},
again by a simpler analysis.

Consider the case that two spatial modes $a,b$ each contain the same
definite number of photons $n$, at the same frequency.  Following
Eq.~(\ref{entstate:circle}), the initial state of the two cavities can
be expressed in the entangled coherent state representation as
\begin{align}
  |\psi\rangle &\equiv|n)_a|n)_b \nonumber \\
  &=\frac{1}{\Pi_n(n)} \int
  \slashb{{\rm d}}\varphi \slashb{{\rm d}}\varphi'
  e^{-{\rm i}n(\varphi+\varphi')}|\sqrt{n}e^{{\rm i}\varphi}\rangle_a
  |\sqrt{n}e^{{\rm i}\varphi'}\rangle_b \, .
\end{align}
The field emission from the cavity is described by a linear output
coupling.  After some time, let $a_1$ ($b_1$) represent the
extracavity output fields and $a_2$ ($b_2$) represent the intracavity
fields; see Fig.~\ref{fig:FockInterference}.  The extracavity modes
$a_1,b_1$ now contain some fraction $\epsilon$ of the total light in
the mode.  The state of the two spatial modes is
\begin{multline}
  |\psi\rangle = \frac{1}{\Pi_n(n)}\int \slashb{{\rm d}}\varphi
  \slashb{{\rm d}}\varphi'
  e^{-{\rm i}n(\varphi+\varphi')}|\sqrt{\epsilon
  n}e^{{\rm i}\varphi}\rangle_{a_1}|\sqrt{\epsilon
  n}e^{{\rm i}\varphi'}\rangle_{b_1}\\
  \otimes|\sqrt{(1-\epsilon)n}e^{{\rm
  i}\varphi}\rangle_{a_2}|\sqrt{(1-\epsilon)n}e^{{\rm i}\varphi'}
  \rangle_{b_2} \, .
\end{multline}
Note that the linear coupling does not maintain a Fock state in the
cavity: an indefinite number of photons are leaked out, determined by
the coupling parameter $\epsilon$.  The output modes $a_1,b_1$ are
then combined on the beam splitter, and the resulting state
$|\psi'\rangle \equiv U(\pi/4,0)|\psi\rangle$ is
\begin{multline}
  |\psi'\rangle
  =\frac{1}{\Pi_n(n)}\int \slashb{{\rm d}}\varphi
  \slashb{{\rm d}}\varphi'
  e^{-{\rm i}n(\varphi+\varphi')} \\
  \times |\sqrt{\tfrac{1}{2}\epsilon
  n}(e^{{\rm i}\varphi}+e^{{\rm
  i}\varphi'})\rangle_{a_1}|\sqrt{\tfrac{1}{2}\epsilon
  n}(-e^{{\rm i}\varphi}+e^{{\rm i}\varphi'})\rangle_{b_1}\\
  \otimes|\sqrt{
  (1-\epsilon)n}e^{{\rm i}\varphi}\rangle_{a_2}|\sqrt{
  (1-\epsilon)n}e^{{\rm i}\varphi'}\rangle_{b_2} \,.
\end{multline}

After the beam splitter, photodetection is performed on each mode.
Consider the result where $A$ photons are detected in mode $a_1$ and
$B$ photons are detected in mode $b_1$ after the beam splitter.  The
consequence of this measurement is that the state $|\psi'\rangle$ is
collapsed to $|\psi''\rangle \propto \left[( A|(
  B|\right]|\psi'\rangle$, which we write (ignoring normalization now)
\begin{multline}
  \label{eq:CollapsedCavityState}
  |\psi''\rangle \propto\!\!\int\!\! \slashb{{\rm d}}\varphi \slashb{{\rm
      d}}\varphi' e^{-{\rm
      i}n(\varphi+\varphi')}
  C_{A,B}(\varphi,\varphi')\\
  \times|\sqrt{
  (1-\epsilon)n}e^{{\rm i}\varphi}\rangle_{a_2}|\sqrt{
  (1-\epsilon)n}e^{{\rm i}\varphi'}\rangle_{b_2} \, ,
\end{multline}
where
\begin{equation}
  \label{C}
  C_{A,B}(\varphi,\varphi') =( A|\sqrt{\tfrac{\epsilon
  n}{2}}(e^{{\rm i}\varphi}+e^{{\rm i}\varphi'})\rangle(
  B|\sqrt{\tfrac{\epsilon n}{2}}(-e^{{\rm i}\varphi}+e^{{\rm
  i}\varphi'})\rangle \, .
\end{equation}
The effect of the collapse is that the distribution over
$\varphi,\varphi'$ is no longer uniform, as captured by the function
$C_{A,B}(\varphi,\varphi')$.  Ignoring factors that are independent of
$\varphi,\varphi'$ and which are removed by normalization, we have
\begin{equation}
  \label{eq:MagC}
  C_{A,B}(\varphi,\varphi') \propto
  e^{-i(A+B)(\varphi+\varphi')/2} |\cos\Delta|^A|\sin\Delta|^B \, ,
\end{equation}
where $\Delta\equiv(\varphi-\varphi')/2$, and where we have used the
expansion~(\ref{coherentstate}) of coherent states in terms of number
states. Note that the presence of the factors $e^{-{\rm
    i}n(\varphi+\varphi')}$ and $e^{-i(A+B)(\varphi+\varphi')/2} $
ensure that (\ref{eq:CollapsedCavityState}) is still a state of
definite photon number.  Moreover, it is a highly entangled state, and
as mentioned above such entanglement will be highly robust - to
destroy this entanglement requires a violation of the PNSSR. The
robustness of such entanglement was first noted and investigated
numerically by M{\o}lmer~\cite{Mol97b}.

\begin{figure}
  \includegraphics*[width=3.25in,keepaspectratio]{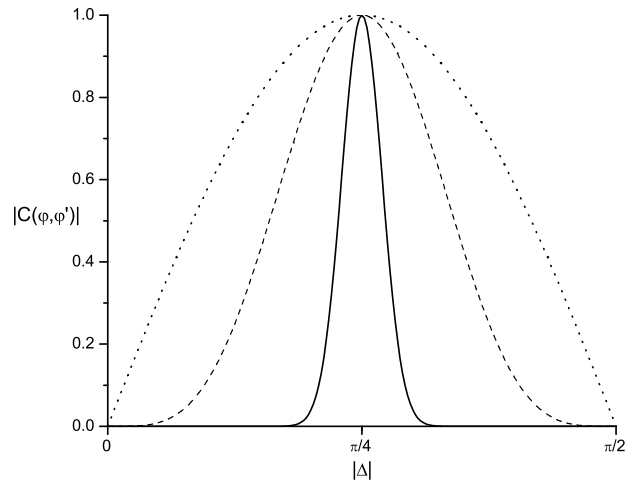}
  \caption{The magnitude $|C_{A,B}(\varphi,\varphi')|$ of the function
    in Eq.~(\ref{eq:CalcWidth}), normalized to have unit magnitude at
    its peak, is plotted as a function of $|\Delta| = |\varphi -
    \varphi'|/2$ for equal photocounts $A=B=1$ (dotted line), $A=B=4$
    (dashed line) and $A=B=64$ (solid line).}
  \label{fig:Cplot}
\end{figure}

The distribution $|C_{A,B}(\varphi,\varphi')|$ is peaked at two
values:
\begin{equation}
  \label{eq:PeakDelta}
  \bar{\Delta}=\pm \arctan(\sqrt{B/A}) \, ,
\end{equation}
within the range $[-\pi/2,\pi/2]$.  Thus, photodetection collapses the
joint state of the cavities into one with correlations in the phase.
Moreover, the width of the distribution over $\Delta$ at each peak
becomes narrower the greater the total number of photons $N=A+B$
detected.  In terms of the difference $\Delta - \bar{\Delta}$ from
each of the maximum values, the relation~\cite{Row01}
\begin{equation}
  \label{WidthRelations}
  |\cos \Delta|^A|\sin \Delta|^B \simeq \sqrt{\frac{A^A B^B}{N^N}}
   \bigl[\cos(\Delta - \bar{\Delta})\bigr]^{2N} \, ,
\end{equation}
valid for large $N$, gives an expansion for~(\ref{eq:MagC}) in terms
of this difference for large $N$ as
\begin{align}
  \label{eq:CalcWidth}
  |C_{A,B}(\varphi,\varphi')| &\propto \bigl[\cos(\Delta -
   \bar{\Delta})\bigr]^{2N} \nonumber \\
  &\propto \exp\bigl(-\tfrac{1}{4}N(\Delta -
   \bar{\Delta})^2\Bigr) \, .
\end{align}
For large $N$, the distribution approaches a Gaussian with standard
deviation proportional to $1/\sqrt{N}$.  Fig.~\ref{fig:Cplot} gives a
plot of the magnitude $|C_{A,B}(\varphi,\varphi')|$ for a specific
ratio $B/A=1$, for various total photon counts $N$.

In the limit $N \to \infty$, the distribution $C(\varphi,\varphi')$
approaches a sum of two delta functions centred at $\pm\bar{\Delta}$.
(The fact that this photodetection measurement only determines a phase
difference between the cavities and does not determine \emph{which}
cavity has the advanced phase results in two peaks rather than one.)
Thus, as a larger number of photons are detected, the state of the
modes $a_2,b_2$ given by~(\ref{eq:CollapsedCavityState}) becomes
closer and closer to a superposition over coherent states with a fixed
relative phase; they become ``phase locked''.  As such, scanning
across a phase shift introduced between the two modes $a_2,b_2$
results in a standard interference pattern, such as is normally
attributed to arising from the interference of two coherent states.

\section{Homodyne detection}
\label{sec:homodyne}

Homodyne detection involves the mixing of a signal field state with a
coherent local oscillator field (typically assumed to be in an
independent coherent state) at a beam
splitter~\cite{Yue78,Yue79,Yue83}, with photodetection at the output
modes.  In balanced homodyne detection~\cite{Yue83}, a 50/50 beam
splitter is employed.  The difference photocurrent for the two
photodetectors is measured and used to infer quadrature phase
statistics for the signal field. By varying the phase $\theta$ of the
local oscillator, homodyne detection over the full set of relative
phases between the signal field and the local oscillator can be
obtained; from these data, the density matrix for the signal field can
be inferred.

It is clear that in the standard description of homodyne detection the
local oscillator provides an absolute phase reference, yet our
preceding analyses make it clear that such a phase reference is not
available in quantum optics.  Although the theory of homodyne
detection is well understood~\cite{Yue78,Yue79}, the interpretation is
predicated on the assumption that the local oscillator is
independently preparable with reasonably definite overall phase. Our
objective in this section is to show that homodyne detection is just
as effective for number state sources, albeit with the restriction
(always employed in practice in quantum optics) that the signal field
and local oscillator are derived from the same source. This important
requirement is simply that the signal and local oscillator have a
localized phase \emph{difference}; describing number state sources in
the entangled coherent state representation clearly reveals how this
requirement is satisfied.

\begin{figure}
  \includegraphics*[width=3.25in,keepaspectratio]{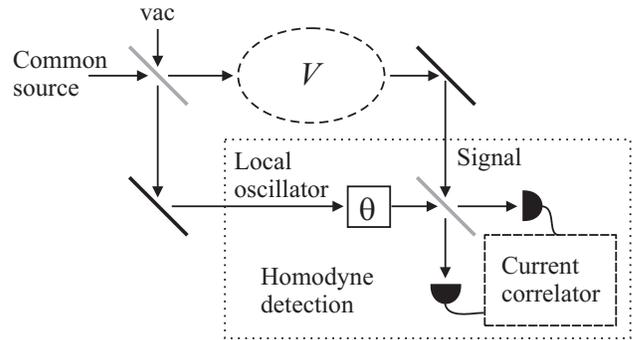}
  \caption{Schematic of a homodyne scheme involving a common source.
    This source is split into a local oscillator and a pre-signal
    field; the pre-signal undergoes a unitary transformation $V$ to
    give the signal field.  All components inside the dashed box
    represent a homodyne detection scheme.}
  \label{fig:Homodyne}
\end{figure}

A schematic for homodyne detection utilizing a common source is given
in Fig.~\ref{fig:Homodyne}.
Consider the case where the common source is a number state $|n)$,
which is mixed with the vacuum via a linear coupler to yield a
pre-signal field and a ``local oscillator''.  The linear mode coupler
output for a beam splitter with choice of relative phase $\phi=-\pi/2$
and reflectivity $r=\cos\theta$ (typically chosen to be near unity,
making the local oscillator strong compared to the pre-signal) is
given by~(\ref{output:n}) as
\begin{multline}
\label{output:first}
  U(\theta,-\pi/2)|n,0)
   =\frac{1}{\sqrt{\Pi_n(n)}}
   \int\slashb{\text{d}}\varphi\text{e}^{-\text{i}n\varphi}
       \\ \times
   |\sqrt{n}\cos\theta\text{e}^{\text{i}\varphi},
   \text{i}\sqrt{n}\sin\theta\text{e}^{\text{i}\varphi}\rangle
   \,.
\end{multline}
The first output mode is subjected to a unitary transformation $V$;
the resulting state is
\begin{multline}
  \label{state:V}
  (V\otimes\openone)U(\theta,-\pi/2)|n,0)
            \\
  =\frac{1}{\Pi_n(n)} \int\slashb{\text{d}}\varphi
        \text{e}^{-\text{i}n\varphi}\left(V
        |\sqrt{n}\cos\theta\text{e}^{\text{i}\varphi}\rangle\right)
        |\text{i}
            \sqrt{n}\sin\theta\text{e}^{\text{i}\varphi}\rangle
            \, .
\end{multline}
The validity and convenience of assuming an independent local
oscillator in a coherent state is made evident by the above equation.
If the source were a coherent state, the pre-signal and local
oscillator are in a product state and can be considered independent.
With the number state approach, the local oscillator is not
independent but is rather is entangled with the source of the signal
state.  This approach reveals that the nature of homodyne detection is
interferometric: it can be used to characterize a \emph{process}
(given by the unitary $V$ in this case) rather than a state.  In
particular, reconstruction of the state of the signal mode through
optical homodyne tomography~\cite{Smi93} relies on the belief that the
local oscillator is in a coherent state.  Our analysis reveals that
this belief is unfounded; however, such tomography can be used more
appropriately to reconstruct information about the process regardless
of the nature of the common source.

Of course the above analysis is somewhat simplified, and more general
signal field states can certainly be considered - such as homodyne
detection of one mode of a two-mode state, decoherence and losses
included in the transformation of the signal mode, entanglement with
ancilla modes and so on.  However, the conclusions for these cases
remain unaffected.

In summary, quantum optics sources satisfy the invariance
condition (\ref{rho:invariance}), and, therefore, independent
local oscillators with specified optical phase are not available.
The reason that we assume independent local oscillators is that
the local oscillator and the signal field are phase-locked, for
example by originating from the same coherent source.  A
decomposition of the density operator in the coherent state basis
makes this clear but has also led to the misconception that
coherent states are the ``actual physical'' states. Here we have
shown how the same result occurs by assuming that the source
produces number states instead of coherent states and demonstrated
that the entangled coherent state representation yields, in a
transparent way, an interpretation of homodyne detection as taking
place on an entanglement of product states, one for the signal and
the other for the local oscillator, with the entanglement being
over the optical phase.

\section{Squeezed light}
\label{sec:squeezed}

The generation of two-mode squeezed light is described by the
interaction Hamiltonian~\cite{Wal94}
\begin{equation}
\label{H:quantumpump}
     \hat{H}_\text{sq}(\zeta) = \text{i}(
     \zeta^* \hat{c}^\dagger\hat{a}\hat{b}
     -\zeta\hat{c}\hat{a}^\dagger\hat{b}^\dagger),
\end{equation}
with $\hat{c}$ the annihilation operator for the pump field, $\hat{a}$
the annihilation operator for the signal field and $\hat{b}$ the
annihilation operator for the idler field.  One pump photon is
annihilated via this process to produce a pair of signal and idler
photons that are correlated in momentum, energy, time of creation and
joint quadrature phase measurements.  The unitary evolution generated
by the squeezing Hamiltonian is given by
\begin{equation}
\label{U:squeezing}
U_\text{sq}(\zeta t)
        =\exp(\zeta^* t\hat{c}^\dagger\hat{a}\hat{b}
        - \zeta t\hat{c}\hat{a}^\dagger\hat{b}^\dagger),
\end{equation}
with $t$ the time of evolution.  Calculations with these three-mode
operators are cumbersome and are generally done either in the
low-$(\zeta t)$ limit or by replacing the pump field annihilation
operator $\hat{c}$ by a c-number.

This c-number replacement is employed in investigating squeezed
light, such as that generated by a second-order nonlinear optical
process in the below-threshold regime.  Ideal two-mode squeezing
is then obtained if the pump field is treated as a classical
coherent pump field with a definite phase.  In this case, we
replace $\hat c$ by $\gamma$, with arg$(\gamma)$ the phase of the
pump field, and let $\chi=\zeta\gamma^*t$.  Then the idealized
squeezing unitary evolution is given by
\begin{equation}
\label{U:idealsqueezing}
  U_\text{sq}(\chi)
        =\exp( \chi^*\hat{a}\hat{b}
        - \chi\hat{a}^\dagger\hat{b}^\dagger ) \,.
\end{equation}
Thus, by treating the pump as a classical field with a definite phase,
the effect of this transformation on the vacuum state for modes $a$
and $b$ is the two-mode squeezed vacuum
\begin{equation}
  \label{eq:TwoModeSqueezedVac}
  |\chi\rangle_{ab} = U_\text{sq}(\chi)|0,0)_{ab} \, .
\end{equation}

It should be noted that two-mode squeezing, as described by the
unitary evolution operator (\ref{U:idealsqueezing}), can equally well
be generated by two single-mode squeezers mixed at a beam
splitter~\cite{Fur98,Zha03,Bow03}, where the same pump field is used
for both squeezers and has definite phase (that is transferred to the
squeezing orientation); the ideal single-mode squeezing Hamiltonian is
given by
\begin{equation}
    H = \chi\hat{a}^2+\chi^*\hat{a}^{\dagger 2}.
\end{equation}
However, we discuss only the direct generation of two-mode
squeezing; the principles elucidated here apply just as simply to
the case of two-mode squeezing generated by mixing two single-mode
squeezed fields.

Consider now squeezing where the pump field is in a number state
$|n)$. Again, expressing this number state in our coherent state
representation, the squeezing transformation
(\ref{U:idealsqueezing}) gives
\begin{multline}
  \label{eq:SqueezedNumberState}
  U_\text{sq}(\chi) |n)_c |0,0)_{ab} = \frac{1}{\sqrt{\Pi_n(n)}}
        \int\slashb{\text{d}}\varphi
        \text{e}^{-\text{i}n\varphi} \\ \text{e}^{\zeta^* t \hat{c}^\dag
        \hat{a} \hat{b} - \zeta t \hat{c} \hat{a}^\dag \hat{b}^\dag}
        |\sqrt{n}\text{e}^{\text{i}\varphi}\rangle_c |0,0\rangle_{ab}
        \, .
\end{multline}
Care must be taken in making the analog of the classical pump
approximation for a coherent state source.  However, if $n$ is large,
it is valid to replace $\hat{c}$ with the c-number
$\sqrt{n}\text{e}^{\text{i}\varphi}$ inside the integral.  Defining
$\chi(\varphi) = \sqrt{n}\zeta t \text{e}^{\text{i}\varphi}$ and using
Eq.~(\ref{eq:TwoModeSqueezedVac}) yields
\begin{multline}
  \label{eq:NumberSqueezedVac}
  U_\text{sq}(\chi) |n)_c |0,0)_{ab} \\ \simeq \frac{1}{\sqrt{\Pi_n(n)}}
        \int\slashb{\text{d}}\varphi
        \text{e}^{-\text{i}n\varphi}
        |\sqrt{n}\text{e}^{\text{i}\varphi}\rangle_c
        |\chi(\varphi)\rangle_{ab} \, .
\end{multline}
Thus, the modes $a$ and $b$ are in a two-mode squeezed state,
entangled via the phase with the state of the pump.  This state
clearly exhibits the distillable entanglement of van Enk and
Fuchs~\cite{Enk02}: an appropriate measurement on the pump mode will
collapse modes $a$ and $b$ into a two-mode entangled state.  Note,
however, that such a measurement violates the PNSSR.

\section{Conclusions}
\label{sec:conclusions}

The fact that quantum optics operationally obeys a PNSSR ensures that
it is equally valid to treat all sources as either distributions of
number states or coherent states.  Traditionally, the coherent state
approach has been standard due to the ease of calculations. Here, we
have presented a powerful and useful tool to carry out calculations
using number state sources with the ease of coherent states through
the entangled coherent state representation.  We have demonstrated
that, in many standard concepts and experiments in quantum optics
where it appears necessary to employ coherent states, it is equally as
valid to describe them using sources of number states. In addition, we
have shown how to provide a simple analysis of the interference
between two initially independent Fock states of photons.

Considerable debate has occurred over the use of coherent states in
continuous variable quantum teleportation.  In quantum teleportation,
a quantum state can be transmitted by two parties (referred to as
Alice and Bob) who share an entangled resource and a classical
communication channel.  In the standard nomenclature, Alice is the
sender, and she performs a joint measurement on her received quantum
state and her portion of the entanglement resource and sends the
results of her measurement to Bob.  Bob transforms his portion of the
entanglement resource into a replica of the original state based on
the classical information he receives from Alice.

One experimental approach to quantum teleportation has been the
teleportation of coherent states~\cite{Fur98,Zha03,Bow03}.  However,
as we have shown, coherent states are not physical but rather just a
convenient representation.  Moreover, a description involving number
state sources should be equally valid.  The teleportation of coherent
states is thus quite interesting because this interpretation is only
meaningful if the coherent state decomposition of the density matrix
is adopted.  It has been suggested by van Enk and Fuchs~\cite{Enk02}
that acquiring a technology for complete phase measurements could
overcome this hurdle, but as we have discussed, these measurements
would break the PNSSR.  As our results show, it would be equally valid
to carry out the calculations for continuous variable quantum
teleportation for a number state source (using the entangled coherent
state representation).  The result would no longer be interpretable as
a standard quantum teleportation experiment, because the state to be
teleported, the shared squeezed vacuum of Alice and Bob, and the local
oscillators used by Alice and Bob for homodyne measurements,
displacements, and final verification of quantum teleportation are all
entangled via the linear coupling of the common source
field~\cite{Rud01}.

\begin{acknowledgments}
  B.C.S.\ and S.D.B.\ acknowledge support from the Australian Research
  Council and the Australian Department of Education, Science and
  Training IAP Grant to support the European Fifth Framework Project
  QUPRODIS.  T.R.\ is supported by the NSA \& ARO under contract No.
  DAAG55-98-C-0040.  P.L.K.\ acknowledges support from the UK
  Engineering and Physical Sciences Research Council and the European
  Union.  We appreciate useful discussions with Howard Carmichael,
  Robert Spekkens and Tom\'{a}\v{s} Tyc.
\end{acknowledgments}


\begin{thebibliography}{99}

\bibitem {Scu97} M. O. Scully and M. S. Zubairy, \emph{Quantum Optics}
  (Cambridge University Press, Cambridge, 1997).

\bibitem {San02} C. Santori, D. Fattal, J. Vuckovic, G. S. Solomon and
  Y. Yamamoto, Nature (Lond.)  \textbf{419}, 594 (2002)

\bibitem {Var00} B. T. H. Varcoe, S. Brattke, M. Weidinger and H.
  Walther, Nature (Lond.)  \textbf{403}, 6771 (2000).

\bibitem {Gla63} R. J. Glauber, Phys. Rev. \textbf{131}, 2766 (1963).

\bibitem {Mol97} K. M\o lmer, \pra \textbf{55}, 3195 (1997);
  J. Gea-Banacloche, \pra \textbf{58}, 4244 (1998);
  K. M\o lmer, \pra \textbf{58}, 4247 (1998).

\bibitem {Rup95}
  P. A. Ruprecht, M. J. Holland and K. Burnett, \pra \textbf{51}, 4704 (1995);
  S. M. Barnett, K. Burnett and J. A. Vaccaro,
  J. Res. Nat. Inst. Stand. Technol. \textbf{101}, 593 (1996).

\bibitem {Yue78} H.~P.~Yuen and J.~H.~Shapiro,
                in {\em Coherence and Quantum Optics IV}, edited by
                L.~Mandel and E.~Wolf (Plenum, New York, 1978),
                p.~719. 

\bibitem {Fur98} A. Furusawa, J. L. S\o orensen, S. L. Braunstein, C.
  A. Fuchs, H. J. Kimble, and E. S. Polzik, Science \textbf{282}, 706
  (1998).

\bibitem {Bow03} W. P. Bowen, N. Treps, B. C. Buchler, R.
        Schnabel, T. C. Ralph, H.-A. Bachor, T. Symul, and P. K. Lam,
        \pra \textbf{67}, 032302 (2003)

\bibitem {Zha03} T. C. Zhang, K. W. Goh, C. W. Chou, P. Lodahl, and
  H. J. Kimble, \pra \textbf{67}, 033802 (2003).

\bibitem {Smi93} K. Vogel and H. Risken, \pra \textbf{40}, R2847
 (1989); D. T. Smithey, M. Beck, M. G. Raymer and A. Faridani,
  \prl \textbf{70}, 1244 (1993).

\bibitem {Rud01} T. Rudolph and B. C. Sanders, \prl \textbf{87},
  077903 (2001). 

\bibitem {Jav96} J. Javanainen and S. M. Yoo, \prl \textbf{76}, 161
  (1996).

\bibitem {Enk02} S. J. van Enk and Ch. A. Fuchs, \prl
 \textbf{88}, 027902 (2002).

\bibitem {Wis02} H. M. Wiseman, arXiv:quant-ph/0104004;
  J. Mod. Opt. \textbf{50}, 1797 (2003).

\bibitem {Fuj03} M. Fujii, arXiv:quant-ph/0301045;
  arXiv:quant-ph/0304148.

\bibitem {Row99} D. J. Rowe, B. C. Sanders, and H. de Guise, J.
        Math. Phys. \textbf{40}, 3604 (1999).

\bibitem {Row01} D. J. Rowe, H. de Guise,
        and B. C. Sanders, J. Math. Phys. \textbf{42}, 2315 (2001).

\bibitem {Mec87} R. Mecozzi and P. Tombesi, J. Opt. Soc. Am. B
\textbf{4}, 1700 (1987).

\bibitem {San92} B. C. Sanders, \pra \textbf{45}, 6811 (1992);
  \textbf{46}, 2966 (1992).

\bibitem {San95} B. C. Sanders, K. S. Lee, and M. S. Kim,
        \pra \textbf{52}, 735 (1995).

\bibitem {San00} B. C. Sanders and D. A. Rice, \pra \textbf{61}, 013805 (2000);
  X. Wang and B. C. Sanders, \pra \textbf{65}, 012303 (2002).

\bibitem {Mun01} W. J. Munro, G. J. Milburn and B. C. Sanders,
   \pra \textbf{62}, 052108 (2001).

\bibitem {Sud63} E. C. G. Sudarshan, \prl \textbf{10}, 277 (1963).
  
\bibitem {Bar03} S. D. Bartlett and H. M. Wiseman, \prl \textbf{91},
  097903 (2003).

\bibitem {Buz95} V. Bu\v{z}ek and P. L. Knight, Opt. Comm.
  \textbf{81}, 331 (1991); Prog. Opt. \textbf{XXXIV}, 1, E. Wolf, ed.
  (Elsevier, Amsterdam, 1995).

\bibitem {Rud02} T. Rudolph and B. C. Sanders,
  arXiv:quant-ph/0112020.

\bibitem {San99} B. C. Sanders, D. J. Rowe, H. de Guise, and A. Mann,
        J. Phys. A: Math. Gen. \textbf{32}, 7791 (1999).

\bibitem {Wal94} D. F. Walls and G.\ J.\ Milburn, \emph{Quantum Optics}
  (Springer-Verlag, Berlin, 1994).

\bibitem {Pfl67} R. L. Pfleegor and L. Mandel, Phys. Lett. A
  \textbf{24}, 766 (1967); Phys. Rev. \textbf{159}, 1084 (1967);
  J. Opt. Soc. Am. \textbf{58}, 946 (1968).

\bibitem {Mol97b} K. M\o lmer, J. Mod. Opt. \textbf{44}, 1937 (1997).

\bibitem {Yue79} H.~P.~Yuen and J.~H.~Shapiro, IEEE~Trans.~Inf.~Theory
                \textbf{IT-25}, 179 (1979); \textbf{IT-26}, 78 (1980);
                H.~P.~Yuen and J.~H.~Shapiro, IEEE~Trans.~Inf.~Theory
                \textbf{IT-26}, 78 (1980).

\bibitem {Yue83} H. P. Yuen and V. W. S. Chan, Opt.
  Lett. \textbf{8}, 177 (1983).

\end{thebibliography}
\end{document}